\documentclass[
  aps,prr,reprint,longbibliography,   % PRR + include titles
  nofootinbib,preprintnumbers
]{revtex4-2}

\usepackage{amsmath,amssymb,bm,mathrsfs}
\usepackage{graphicx,color}
\usepackage{amsfonts}
\usepackage[export]{adjustbox}
\usepackage{mathtools}
\usepackage{srcltx}
\usepackage{dsfont}
\usepackage{epsfig}
\usepackage{bbold}
\usepackage{rotating}
\usepackage[dvipsnames]{xcolor}
\usepackage[caption=false]{subfig}
\usepackage{xfrac}
\usepackage{multirow}
\usepackage{booktabs}
\usepackage{soul}
\usepackage[inline]{enumitem} % for inline list
\usepackage[english]{babel}

\makeatletter
\@namedef{captionsen}{\captionsenglish}
\@namedef{dateen}{\dateenglish}
\@namedef{extrasen}{\extrasenglish}
\@namedef{noextrasen}{\noextrasenglish}
\makeatother
\usepackage[colorlinks=true
,urlcolor=blue
,anchorcolor=blue
,citecolor=blue
,filecolor=blue
,linkcolor=red
,menucolor=blue
,linktocpage=true
,pdfproducer=medialab
,pdfa=true
]{hyperref}
\usepackage{cleveref}
\usepackage{setspace}
\usepackage[
   justification=justified,
   format=plain]{caption}
\usepackage{booktabs}

\usepackage{tikz}

\definecolor{forestgreen}{RGB}{34,139,34}

\newcounter{qnumber}

\begin{document}

%=============================================================================

\title{Dark plasmas in the nonlinear regime: constraints from particle-in-cell simulations}

\author{William DeRocco}

\affiliation{Maryland Center for Fundamental Physics, University of Maryland, College Park, 4296 Stadium Drive, College Park, MD 20742, USA}
\affiliation{Department of Physics \& Astronomy, The Johns Hopkins University, 3400 N. Charles Street, Baltimore, MD 21218, USA}

\author{Pierce Giffin}

\affiliation{Department of Physics, University of California Santa Cruz, 1156 High St., Santa Cruz, CA 95064, USA}
\affiliation{Santa Cruz Institute for Particle Physics, 1156 High St., Santa Cruz, CA 95064, USA}
%=============================================================================

%-----------------------------------------------------------------------------
\begin{abstract}
If the dark sector possesses long-range self-interactions, these interactions can source dramatic collective instabilities even in astrophysical settings where the collisional mean free path is long. Here, we focus on the specific case of dark matter halos composed of a dark $U(1)$ gauge sector undergoing a dissociative cluster merger. We study this by performing the first dedicated particle-in-cell plasma simulations of interacting dark matter streams, tracking the growth, formation, and saturation of instabilities through both the linear and nonlinear regimes. We find that these instabilities give rise to local (dark) electromagnetic inhomogeneities that serve as scattering sites, inducing an effective dynamic collisional cross-section. Mapping this effective cross-section onto existing results from large-scale simulations of the Bullet Cluster, we extend the limit on the dark charge-to-mass ratio by over ten orders of magnitude. Our results serve as a simple example of the rich phenomenology that may arise in a dark sector with long-range interactions and motivate future dedicated study of such ``dark plasmas.''
\end{abstract}

\maketitle
\section{Introduction}
\label{sec:intro}
There are multiple lines of compelling evidence that indicate the existence of an unknown matter content termed ``dark matter'' (DM)~\cite{Sofue:2000jx,Planck:2018vyg,Massey:2010hh,Primack:2015kpa}, the microphysical nature of which remains unknown. 
Vast theoretical and experimental effort has been devoted to discovering the nature of dark matter, often under the assumption that dark matter consists of a new fundamental particle with a low interaction cross-section with the visible sector. As a result, the phenomenology of dark matter has relied almost exclusively on single-particle interactions. 
This is in direct contrast to the visible sector of the Universe, in which the vast majority of visible matter exists in the form of plasma, the behavior of which is governed not by single-particle interactions but by collective effects mediated by long-range interactions. The most dramatic of these effects arise from  \textit{plasma instabilities}, in which small initial perturbations grow exponentially and produce nonlinear structures even on scales far smaller than the particle scattering mean free path. If, in analogy to the visible sector, self-interactions in the dark sector are dominated by collective effects, then the associated instabilities can lead to observable effects in astrophysical systems.

Despite this, little work has been done on characterizing dark sectors governed by collective interactions. Under the assumption that collective effects would make dark matter effectively collisional, Heinkinheimo et al.~\cite{Heikinheimo:2015kra,Heikinheimo:2017meg,Spethmann:2016glr} performed hydrodynamic simulations of dissociative cluster mergers with an artificial viscosity that was tuned to best reproduce observed lensing maps. Growth rates for common instabilities have been computed in the linear regime for a variety of astrophysical settings~\cite{Lasenby:2020rlf,Cruz:2022otv}. Simulations of interpenetrating electron-positron pair plasmas have also discussed possible implications for the dark sector~\cite{Shukla_2022}, and analytic arguments for the existence of limits on particular models have also appeared in the literature \cite{medvedev2024plasmaconstraintsmillichargeddark}.

Here, we perform dedicated fully-kinetic plasma simulations with astrophysically-motivated parameters to measure the effective self-interaction cross-section of dark matter well into the nonlinear regime. While the model we adopt exhibits the same classes of instability as the Standard Model electromagnetic sector, we for the first time robustly connect the nonlinear behavior of these micro-instabilities to existing large-scale astrophysical simulations of dark matter self-interactions. Our analysis bridges the gap between plasma microphysics and dark matter macrophysics, allowing us to place constraints on dark couplings roughly ten orders of magnitude below existing constraints.

\section{Dark U(1) Plasma}
\label{sec:model}

The Standard Model (SM) electromagnetic sector exhibits a rich phenomenology of plasma behaviors; a dark sector with a massless $U(1)$ mediator is therefore a natural place to look for similar effects. As such, we will take as our fiducial model that of ``dark electromagnetism,'' namely a sector composed of ``dark'' electrons and positrons interacting via a massless $U(1)$ gauge boson. This model has been explored previously in the literature \cite{Ackerman_2009,Feng_2009,Agrawal_2017} and has been found to be able to produce the correct relic abundance of dark matter \cite{Agrawal_2017}. The Lagrangian is given by
\begin{equation}
    \mathcal{L} = \mathcal{L}_\text{SM} - \frac{1}{4}F'_{\mu\nu}F'^{\mu\nu} + \bar{\chi}(\gamma^\mu(i\partial_\mu - q_\chi A_{\mu}')- m_\chi)\chi
\end{equation}
where $\chi$, $\bar{\chi}$ are dark fermion fields with mass $m_\chi$ and charge $q_\chi$ under a new $U(1)$ gauge symmetry mediated by a massless ``dark photon'' with the four-potential $A_\mu'$. Here, $F_{\mu\nu}'=\partial_\mu A_\nu'-\partial_\nu A_\mu'$ is the dark electromagnetic field strength tensor, $\mathcal{L}_\text{SM}$ is the standard model Lagrangian, and $\gamma^\mu$ are the Dirac matrices. While the particle content is similar to the SM electromagnetic sector, the charge $q_\chi$ and mass $m_\chi$ of the fermions are taken to be free parameters. We restrict our attention to the case of a massless mediator and assume throughout this paper that the dark sector is completely decoupled from the Standard Model, i.e., there is no kinetic mixing term $\epsilon F_{\mu\nu} F'^{\mu \nu}$. As discussed in Ref. \cite{Ackerman_2009}, this is a self-consistent choice, as if $\epsilon$ is set to zero at some high scale, $\epsilon = 0$ is preserved under renormalization group evolution.

\begin{figure}[!t]
    \centering
    \includegraphics[width=\textwidth/2]{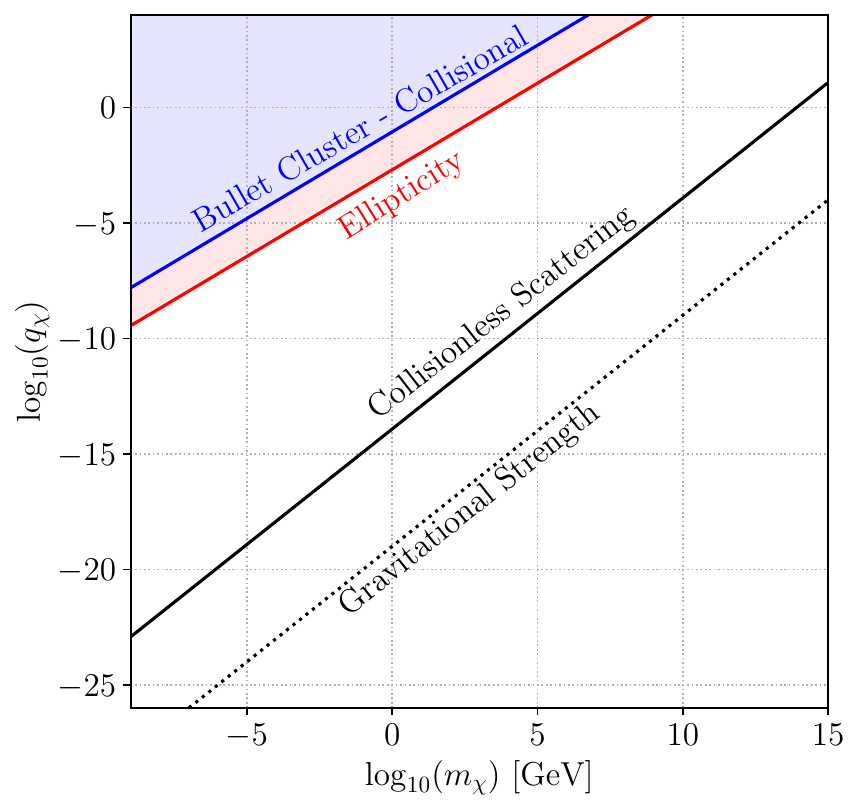}
    \captionsetup{justification=justified}
    \caption{Constraints on dark $U(1)$ electromagnetism in the $q_\chi-m_\chi$ plane. Bullet Cluster constraints from two-to-two scattering are shown in blue. Constraints from the survival of elliptical galaxies are shown in red \cite{Ackerman_2009,Feng_2009,Agrawal_2017}. The results presented in this work place constraints above the solid black line. }
    \label{fig:param}
\end{figure}

Neglecting plasma effects, the strongest existing constraints on this model have been found to arise from the survival of elliptical galaxies on cosmological timescales \cite{Ackerman_2009,Agrawal_2017}. Dissociative cluster mergers also bound the self-interactions of this model, however, when purely considering two-to-two scattering, the associated bounds are several orders of magnitude weaker than the ellipticity bound (see Fig. \ref{fig:param}) \cite{Ackerman_2009,Feng_2009,Agrawal_2017}. 
Despite this, as we show in Sec. \ref{sec:results}, taking into account \textit{collective} effects allows such mergers to constrain couplings as low as ten orders of magnitude below the ellipticity bound (Fig. \ref{fig:param}).

This is because at the microphysical level, the interactions of a dark $U(1)$ sector in the Bullet Cluster can be well-modeled by two counter-streaming beams of net-neutral pair plasma moving at a relative velocity $v_0$. We can confirm that such a plasma approximation is valid by checking that the ``plasma parameter,'' which is the number of particles within a sphere of radius  $\lambda_D$ (the effective screening length in the plasma), is much greater than 1. This screening length is set by the ratio of the particle velocity dispersion $\sigma$ to the plasma frequency $\omega_{\text{pl}} =\sqrt{\frac{q_{\chi}^2 n_\chi}{m_\chi}}$ where $n_\chi$ is the number density of the dark matter particles and we adopt units such that $\hbar=c=\varepsilon_0=\mu_0=1$. In the region of parameter space between existing ellipticity bounds and our constraint, $\lambda_D$ varies between $\sim 10^{-14}-10^{-3}$ kpc for a 1 GeV particle, corresponding to a plasma parameter of $\Lambda=4\pi n_\chi\lambda_D^3\sim 10^{23}-10^{56}\gg 1$, which indicates that the plasma approximation is valid and the system is dominated by collective effects.

As described in detail in the Appendix (see also \cite{Bret_2020,Bret_Deutsch_2005,Bret_Gremillet_Dieckmann_2010,dieckmann_bret_2017,Bret_Haggerty_Narayan_2024,Bret_Narayan_2022,Bret_Stockem_etal_2016,Livadiotis_2019,Plasma_Textbook,Bret_Stockem_Fiúza_Álvaro_Ruyer_Narayan_Silva_2013,NH1,NH2,NH3,1D_Weibel,Medvedev_2005,Ruyer_Fiuza_2018}), this system is subject to two primary instabilities: the electrostatic \textit{longitudinal instability} \cite{TS_Original} and the electromagnetic \textit{Weibel instability} \cite{Weibel_Original}\footnote{Additional electromagnetic oblique modes can occur in asymmetric beam plasmas. For beams that are approximately symmetric, as in this work, their growth rates and saturation mechanisms only vary slightly and can be determined numerically. See \cite{Bret_Gremillet_Dieckmann_2010} for further discussion.}. These instabilities act to slow the relative motion of the two beams while heating each individually, even in a fully collisionless regime \cite{Shukla_2022}.
The longitudinal instability grows at a rate proportional to the plasma frequency which we find to be
\begin{equation}
\label{eq:omegapl}    \omega_{\text{pl}}=0.5\,\text{kyr}^{-1}\left(\frac{q_\chi/m_\chi}{2\times10^{-14}\,\text{GeV}^{-1}}\right)\left(\frac{\rho_{\text{DM}}}{0.035\,\text{GeV}/\text{cm}^3}\right)^{1/2}     
\end{equation}
where $\rho_{\text{DM}}$ is the mass density of the dark matter particles. The Weibel instability grows more slowly, at a rate proportional to
\begin{multline}
\label{eq:omegapl}    \omega_{\text{pl}}=5\,\text{Myr}^{-1}\left(\frac{q_\chi/m_\chi}{2\times10^{-14}\,\text{GeV}^{-1}}\right)\times\\\left(\frac{\rho_{\text{DM}}}{0.035\,\text{GeV}/\text{cm}^3}\right)^{1/2}\left(\frac{v_0}{3000\,\text{km}/\text{s}}\right),
\end{multline}
however it saturates later and at larger electromagnetic field values, hence can play a significant role in the dynamics \cite{Cruz:2022otv}. Prior to saturation, the system can be solved analytically in certain limits, however, once the system saturates, it becomes nonlinear and requires dedicated plasma simulations to model. Note that the Vlasov-Maxwell system that governs the plasma dynamics (see SM) can be recast in a form in which the only unitful quantity is the plasma frequency, hence results for particular values of $(q_\chi, m_\chi)$ can be generalized to other regions in parameter space at which the plasma takes on the same $\omega_{\text{pl}}$.

\section{The Bullet Cluster}
\label{sec:bulletcluster}

The Bullet Cluster (1E 0657-56) \cite{BC1,BC2,BC3,BC4} is a well-known dissociative cluster merger that has been used extensively to constrain dark matter self-interactions \cite{BCDM1,BCDM2,Randall_Markevitch_Clowe_Gonzalez_Bradac_2008,Robertson_2016,Kahlhoefer_Schmidt-Hoberg_Frandsen_Sarkar_2014}. The system consists of a large main galaxy cluster of mass $\approx 10^{15}~M_\odot$ and a smaller subcluster (the ``Bullet'') of mass $\approx 10^{14}~M_\odot$ that merged at a high relative velocity. The primary observable provided by weak lensing maps of the system is on the observed offset between the mass in dark matter enclosed in the central 150 kpc of the Bullet and the shocked Standard Model gas, which trails behind, which
 has been used to constrain the collisional cross-section of self-interacting dark matter with mass $m$ to be $\sigma/m\lesssim 1-4\, \mathrm{cm^2/g}$ \cite{Robertson_2016, Randall_Markevitch_Clowe_Gonzalez_Bradac_2008,Kahlhoefer_Schmidt-Hoberg_Frandsen_Sarkar_2014,TULIN20181}. Though there is tension on the precise value of this constraint, in this paper, we adopt $4\, \mathrm{cm^2/g}$ as a conservative limit.

Existing limits on self-interactions are derived under the assumption of individual particle scattering. However, the Bullet Cluster is also ideal for studying the effects of plasma instabilities on dark sectors charged under a long-range interaction. 
At small scales on which the dark matter density of each cluster varies negligibly, the merger is well-modeled by the two-stream system described in Section \ref{sec:model}. 
Two-stream and Weibel instabilities may arise, leading to the formation of small-scale electromagnetic fields that efficiently scatter DM particles through a wide variety of diffusion mechanisms \cite{Ruyer_Fiuza_2018,Ruyer_Gremillet_Debayle_Bonnaud_2015,Weibel_Bohm,Kaufman_1990,Takabe_2023,Vanthieghem_Lemoine_Gremillet_2018,Chang_2008,McFarland_Wong_2001,McFarland_Wong_2001.2,Yoon_2017} inducing an effective collisional cross-section despite the plasma being largely collisionless. We can compute this effective collisional cross-section via dedicated simulations of the formation and saturation of plasma microinstabilities, as discussed in the following Section. Having extracted these collision rates from our simulations, we can rescale the existing constraints on two-to-two scattering to place constraints on collective interactions as well.\footnote{Note that in this work, we do not explore the formation or stability of dark matter halos, instead assuming that for very weak couplings, the standard picture of structure formation is not appreciably altered. If, however, this is not the case, then stronger constraints may arise from the existence of dark matter halos themselves. We leave a thorough exploration of this to future work.}

\section{Simulations}
\label{sec:simulations}

We simulate two counter-streaming beams of net neutral plasma using \texttt{Smilei}, an open-source particle-in-cell simulation framework \cite{smilei}. To connect our results to the Bullet Cluster, we adopt values taken to be consistent with the analytic Hernquist profiles of the merging clusters provided in \cite{Robertson_2016}, taking $r=150\, \mathrm{kpc}$ for both clusters, corresponding to the maximal radius at which dark matter self-interactions affect observations by weak lensing (see Sec. \ref{sec:bulletcluster}) \cite{Randall_Markevitch_Clowe_Gonzalez_Bradac_2008}. This choice yields velocity dispersions of $\sigma_{\mathrm{Main}} = 1080$ km/s for 
the Main cluster and $\sigma_{\mathrm{Bullet}} = 630\, \mathrm{km/s} $ in the Bullet sub-cluster 
and a relative number density of $n_\text{Main}/n_\text{Bullet}=1.91$. Since both higher temperatures and larger density ratios tend to inhibit instability growth, the values we have selected are conservative. To test this hypothesis, we performed two additional simulations, adopting the parameters that the authors of \cite{Robertson_2016} used to explore the sensitivity of their simulations to initial conditions. Specifically, we increased the mass density of the main cluster to correspond to a halo with concentration parameter $c=1.94$, rather than $c=3$, yielding a density ratio of 1.19 (Table \ref{table:runs}, Simulation R3). We also varied the mass of the Bullet sub-cluster, increasing it by a factor of two in keeping with the mass suggested by strong lensing observations \cite{Robertson_2016}. This yielded a density ratio of 0.96 (Simulation $R4$ in Table \ref{table:runs}.)

The beams are initialized with a relative bulk velocity of  $v_0=3000\, \mathrm{km/s}$, which is consistent with the findings of \cite{Springel_Farrar_2007} that the DM relative velocity may be much less than the observed gas shock velocity ($\approx 4700$ km/s). Furthermore, we note that as larger relative velocities only serve to enhance the microinstabilities, this choice of a low $v_0$ is also conservative. We demonstrate this quantitatively by performing an additional simulation in which $v_0$ is increased to 4000 km/s, which we find significantly decreases the time to saturation. (See Table \ref{table:runs}, R2.)
All other parameters can be absorbed into the plasma frequency as discussed in Section \ref{sec:model}. Our choice of parameters places the system in the ``warm'' regime where the thermal velocity of each stream is comparable to the streams' relative velocity. Warm, non-relativistic pair plasmas are rare in the Standard Model, hence their evolution is less well-studied than the cold and hot regimes, making dedicated simulations even more necessary.

Our simulation consists of a 2D3V\footnote{For a discussion of how this choice affects our results, see Appendix \ref{app:3d}.} Cartesian geometry with periodic boundary conditions. Each simulation was run for $10^3\,\omega_{pl,B}^{-1}$, where $\omega_{pl,B}$ is the plasma frequency corresponding to the Bullet sub-cluster. Two species were initialized, $\chi$ and $\Bar{\chi}$, with equal and opposite charge-to-mass ratios and relative bulk velocity such that the simulation is in the center of momentum frame. See Appendix \ref{app:simulation} for further discussion of simulation setup.

\section{Results}
\label{sec:results}

Throughout the simulation, we record the trajectories of $10^4$ test particles. At regular intervals, we calculate for each particle
\begin{equation}
\label{eq:scatter}
    \Delta v_t = \frac{\vec{v_0}\cdot\vec{v_t}}{|\vec{v_0}|^2}
\end{equation}
where $\vec{v_0}$ is the particle's velocity vector at the start of the simulation and $\vec{v_t}$ is the particle's velocity vector at time $t$. If $\Delta v_t$ changes by a factor of $e$, we then consider the particle to have undergone a hard scatter, similar to the studies performed in \cite{Zhou_2024,Riquelme_2016,Bott2025ThermodynamicsAC}. We can then determine the \textit{two-to-two} scattering cross-section that this fraction of scattered particles would correspond to through \cite{AstrophysicalTests}
\begin{equation}
\label{eq:xsec}
    \frac{\sigma}{m}=-\Sigma^{-1} \log(1-p),
\end{equation}
where $p$ is the fraction of particles originating from the Bullet sub-cluster that have scattered and $\Sigma\sim 0.33\,\mathrm{g/cm^2}$ is the projected mass surface density of the main cluster at a radius of 150 kpc \cite{Robertson_2016}. This relation allows us to connect our results to existing limits on DM self-interactions from large-scale simulations of the Bullet Cluster.
\begin{figure}[!t]
    \centering
    \includegraphics[width=\textwidth/2]{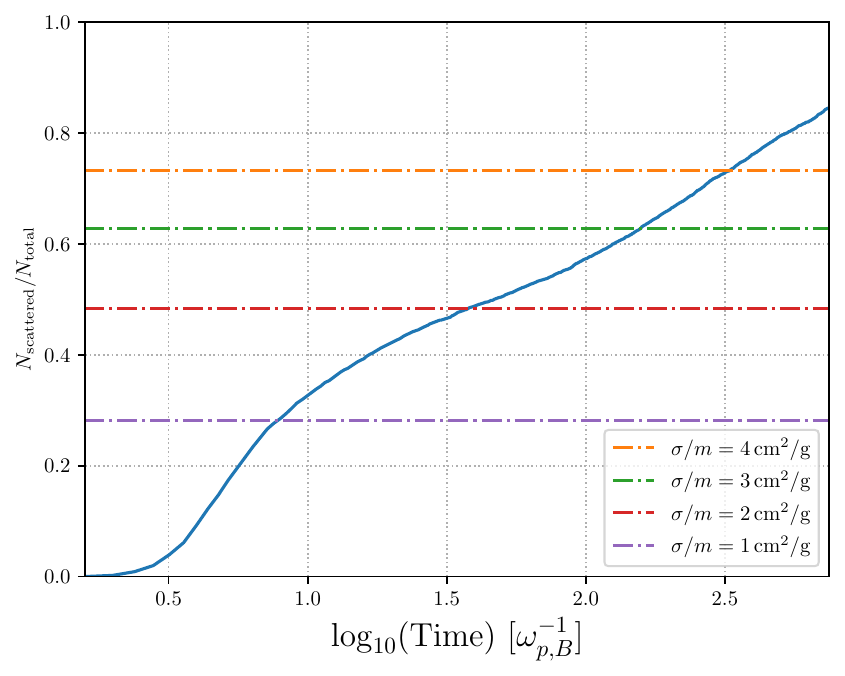}
    \captionsetup{justification=justified}
    \caption{Fraction of tracked macroparticles that have undergone an $\mathcal{O}(1)$ momentum change as a function of time (cf. Eq. \ref{eq:scatter}). The dash-dotted lines show the fraction of particles that existing simulations show undergo \textit{two-to-two} scattering events during the merger for varying cross-sections \cite{AstrophysicalTests}. At 330 $\omega_{pl,B}^{-1}$, 73\% of particles have been scattered, corresponding to an effective cross-section of $4\,\text{cm}^2/\text{g}$.} 
    \label{fig:Scatter}
\end{figure}

Fig. \ref{fig:Scatter} shows a cumulative count of the percentage of test particles that have undergone a momentum change equivalent to a hard scatter as a function of time. Even though the plasma is collisionless, we find that particles have experienced a significant change in momentum mimicking that of a hard collisional scattering process. Previous studies agree that a collisional cross-section of $\sigma/m\sim 4 \,\mathrm{cm^2/g}$ is ruled out by the centroid offsets of the Bullet Cluster's dark matter halo and galaxies \cite{Robertson_2016, Randall_Markevitch_Clowe_Gonzalez_Bradac_2008,Kahlhoefer_Schmidt-Hoberg_Frandsen_Sarkar_2014,TULIN20181}. By Eq. \ref{eq:xsec}, this corresponds to 73.3\% of particles experiencing a significant change in momentum. We therefore place a constraint on all charge-to-mass ratios that effectively scatter 73.3\% of particles in less than $1\%$ of the Bullet Cluster crossing time, where we have chosen 1\% of crossing as a conservative scale over which the halo density varies negligibly ($\Delta \rho / \rho < 5\%$) hence our simulations remain an appropriate treatment of the system. We estimate the crossing time $t_\text{cross}$ as twice the time it takes for the Hernquist mass distributions described in Sec. \ref{sec:bulletcluster} to move from a centroid separation of 300 kpc to 0 kpc assuming the halos were infalling from rest at infinite separation and that the infall does not significantly disrupt the halo shape. This calculation yields $t_\text{cross} = 1.07\times 10^8$ yr.
This leads to a constraint on the dark charge-to-mass ratio of
\begin{equation}
\label{eq:constraint}
    q_\chi < 1.2\times10^{-14} \left(\frac{m_\chi}{\text{GeV}}\right).
\end{equation}

\begin{figure*}[htbp]
    \centering
        \includegraphics[width=\linewidth]{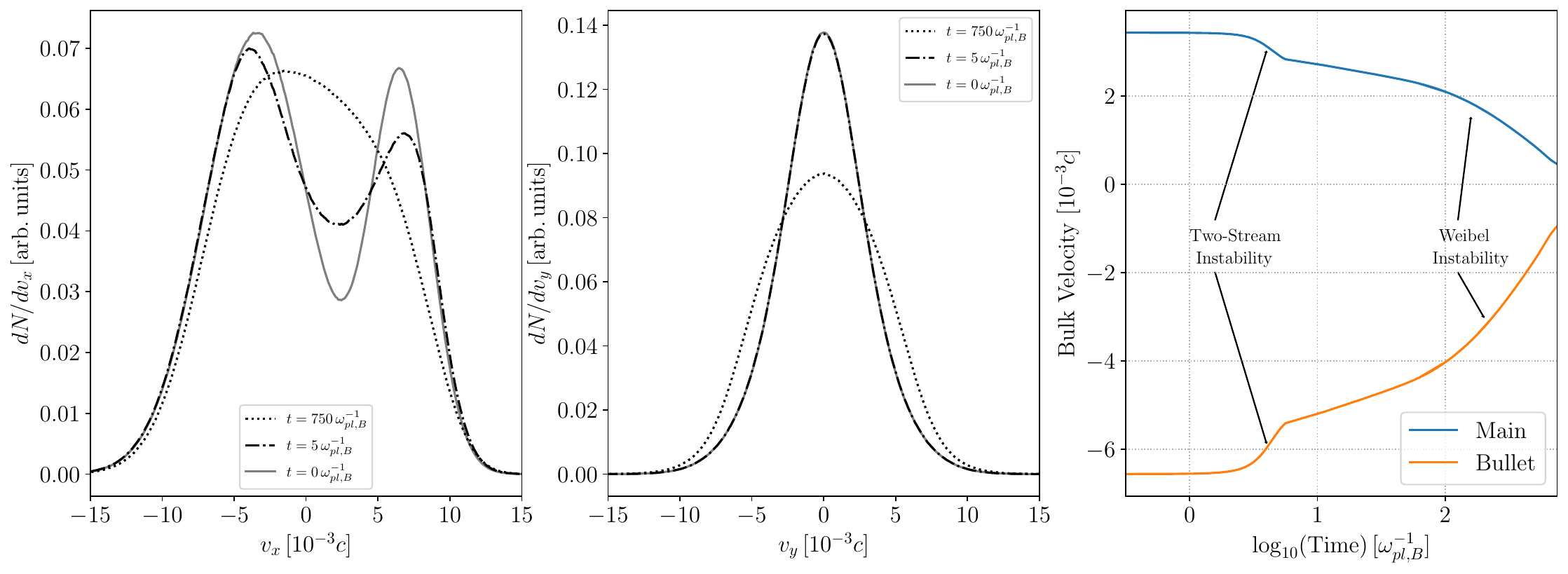}
    \caption{Left: Center of momentum velocity distribution of particles in the longitudinal direction at $t=0\,\omega_{pl,B}^{-1}$ (solid line), $t=5\,\omega_{pl,B}^{-1}$ (dot-dashed line), and $t=750\,\omega_{pl,B}^{-1}$ (dotted line). Middle: Center of momentum velocity distribution of particles in the transverse direction at $t=0\,\omega_{pl,B}^{-1}$ (solid line), $t=5\,\omega_{pl,B}^{-1}$ (dot-dashed line), and $t=750\,\omega_{pl,B}^{-1}$ (dotted line). Right: Bulk velocity of particles in the longitudinal direction for particles originating in the Main sub-cluster beam (blue) and Bullet sub-cluster beam (orange).}
    \label{fig:velocities}
\end{figure*}

This corresponds to a length scale of the fastest growing modes of $\lambda_\text{TS}=0.05$ kpc for the two-stream instability and $\lambda_\text{W}=6.5$ kpc for the Weibel instability. We further verify our results by explicitly computing the slowdown of the bulk velocity of the counter-streaming beams as a function of time. We show the evolution of the velocity distributions in Fig. \ref{fig:velocities}, where the left-hand (center) panel shows the velocity distribution in the longitudinal (transverse) direction at three different times and the right-hand panel shows the bulk velocities of the two beams as a function of time. We find that the bulk velocity decreases by roughly an order of magnitude by the time of Weibel saturation, with the dominant slowdown being due to diffusion of particles by late-time Weibel modes. By the end of our simulation, the two beams have effectively isotropized in the center-of-momentum frame, as can be seen in the dotted distribution corresponding to $t=750\,\omega_{pl,B}^{-1}$ in the left and center panels.
We find these results to be consistent with the observed slowdown found in other similar simulations of nonrelativistic pair plasmas with $v_0\sim0.01c$ and length scales of $\sim 10 c/\omega_{pl}$. The simulation parameters of $R_3$ in \cite{Shukla_2022} match closest to those of our fiducial simulation. After 250 $\omega_{pl}^{-1}$, the authors found that the relative bulk velocity had reduced to $0.442\,v_0$. From our simulation, we find that after the same amount of time, the relative bulk velocity had decreased to $0.444\,v_0$, which we find to be in good agreement.

% \subsection{Additional simulations}
% \label{sec:additionalsimresults}

As discussed in Sec. \ref{sec:simulations}, we performed three additional simulations with varying initial conditions to test the robustness of our results. We varied the relative velocity of the halos, as well as their density ratios, in keeping with the analysis in \cite{Robertson_2016}. The parameter choices and resulting time of scattering 73.3\% of particles, $t_{73.3}$, and charge-to-mass ratio constraint for each simulation are shown in Table \ref{table:runs}. We find that simulation $R4$ produces the most conservative constraint on the dark charge-to-mass ratio. Though the initial conditions chosen for $R4$ are less observationally motivated than those that produced our fiducial limit (Eq. \ref{eq:constraint}), in the interest of being conservative, we adopt 
\begin{equation}
\label{eq:constraint2}
    q_\chi < 2.0\times10^{-14} \left(\frac{m_\chi}{\text{GeV}}\right).
\end{equation}
as our final constraint, corresponding to the bound set by simulation R4.

This constraint is shown in Fig. \ref{fig:param} as a solid black line. We additionally show with the effective charge-to-mass ratio of the gravitational force 
\begin{equation}
    q_G=\sqrt{4\pi G}\left(\frac{m_\chi}{\text{GeV}}\right)
\end{equation}
with a dotted black line for comparison. The proximity of this curve demonstrates that we are probing exceptionally weak couplings; however, as our constraint is still several orders of magnitude above this line, gravitational forces should have negligible effects on the evolution of plasma instabilities in the region of parameter space we constrain. Existing constraints from the ellipticity of dwarf galaxies \cite{Ackerman_2009,Agrawal_2017} (red) and two-to-two scattering limits from the Bullet Cluster (blue) are shown as well. Even for our conservative choice of parameters, our results extend existing limits by over ten orders of magnitude.

\begin{table*}[t]
    \centering
    \begin{tabular}{|c|c|c|c|c||c|c|}
        \hline
        Run & $v_0 \,\left[\mathrm{km/s}\right]$ & $n_\mathrm{Main}/n_\mathrm{Bullet}$ & $\sigma_{\mathrm{Main}}\,\left[\mathrm{km/s}\right]$ & $\sigma_{\mathrm{Bullet}}\,\left[\mathrm{km/s}\right]$  &$t_{73.3}\,\left[\omega_{pl,B_F}^{-1}\right]$& $ q_\chi/m_\chi \,\left[\mathrm{GeV}^{-1}\right]$\\
        \hline
        Fiducial & 3000 & 1.91 & 1080 & 630& 330 & $1.2\times 10^{-14}$\\
        \hline
        R2 & 4000 & 1.91 & 1080 & 630& 20 &$1.0\times 10^{-15}$\\
        \hline
        R3 & 3000 & 1.19 & 900 & 630& 440 & $1.6\times 10^{-14}$\\
        \hline
        R4 & 3000 & 0.96 & 1080 & 880 & 540 & $2.0 \times 10^{-14}$\\
        \hline
    \end{tabular}
    \caption{Initial conditions (left four columns) and resulting constraints (right two columns) from different simulations. R2 utilizes a less conservative estimate of the relative bulk velocity $v_0$ between the two sub-clusters. R3 (R4) utilizes an alternative mass density of the Main (Bullet) sub-cluster. Here, $\omega_{pl,B_F}$ is the plasma frequency of the Bullet sub-cluster from the fiducial simulation. See Sec. \ref{sec:simulations} and \ref{sec:results} for definitions of each parameter.}
    \label{table:runs}
\end{table*}

\section{Conclusions}
\label{sec:conc}

In this work, we have shown that long-range collective effects can have a dramatic impact on the large-scale behavior of dark matter in dissociative cluster mergers. We have performed dedicated particle-in-cell simulations of counter-propagating beams of dark plasma to characterize the growth rate, saturation, and late-time behavior of the beams. We find that instabilities produce electromagnetic inhomogeneities that act as scattering sites for dark matter. This process leads to an effective collisionality in the dark sector, which we then connect to existing limits on DM self-interactions derived from simulations of the Bullet Cluster. The associated results allow us to place new constraints at couplings orders of magnitude below existing limits.

There are many opportunities for future work.
Here, we restricted ourselves to a particular model of the dark sector, namely DM charged under a new dark $U(1)$ gauge symmetry. Other well-motivated models of DM exist (e.g. millicharged DM \cite{Lasenby:2020rlf,Cruz:2022otv,Tongyan_Plasma} and axion DM \cite{Axion_Plasma1,An_Chen_Liu_Sheng_Zhang_2024}) that can also exhibit similar types of instabilities. Additionally, plasma instabilities may affect ``dark atomic'' sectors in which there are both high-mass ``dark protons'' and low-mass ``dark electrons''~\cite{Fan:2013yva,Roy:2023zar,Roy:2024bcu}. 

Furthermore, it is important to note that the constraint presented in this work is likely to be very conservative. Throughout this work, parameters such as the DM thermal velocity, relative beam velocity, and constrained collisional cross-section are all taken at conservative values. Additionally, post-saturation plasma effects are likely to lead to further turbulence and disruption of the halos \cite{Ruyer_Fiuza_2018,Ruyer_Gremillet_Debayle_Bonnaud_2015,Weibel_Bohm,Kaufman_1990,Takabe_2023,Vanthieghem_Lemoine_Gremillet_2018,Chang_2008,McFarland_Wong_2001,McFarland_Wong_2001.2,Yoon_2017}; a dedicated study of these effects could place a more stringent constraint. A more sophisticated macroscopic simulation of the Bullet Cluster system is likely to yield constraints at lower charge-to-mass ratios, and would be a topic well-suited to future exploration.

Our results have only scratched the surface of the rich phenomenology that may exist in the dark sector; if it is anything like its visible counterpart, then it is time to move beyond the interactions of individual particles and instead embrace the collective.

% -----------------------------------------------------------------------------
\vspace{0.2cm}
\noindent {\it Acknowledgements.}
%-----------------------------------------------------------------------------
PG and WD wish to thank Robert Lasenby for useful input during the early stages of this project, as well as Sasha Philippov, Antoine Bret, Brant Robertson, Bart Ripperda, David Curtin, Rouven Essig, Stefano Profumo, Ani Prabhu, Akaxia Cruz, Fr\'ed\'eric P\'erez, Frederico Fi\'uza, and many others for insightful commentary and suggestions throughout the preparation of this manuscript. The authors would also like to thank the anonymous referee for suggesting running more simulations. PG would also like to thank the Smilei community for their interest and assistance in running the simulations. We acknowledge use of the LUX supercomputer at UC Santa Cruz, funded by NSF MRI grant AST 1828315. The authors acknowledge the support of DOE grant No. DE-SC0010107.
\pagebreak

\bibliography{references.bib}

\appendix

\section{Plasma review: Linear and non-linear regimes}
\label{app:plasma}

%\pg{Not sure how you want to organize section labels, but this seems to not be what we want}

Plasmas are quasineutral mixtures of charged particles in which oppositely charged particles are not bound to each other. 
When particle collisions can be neglected, the evolution of the plasma is governed by the Vlasov-Maxwell equations \cite{Plasma_Textbook},
\begin{equation}
    \frac{\partial f_s}{\partial t}+\vec{v}\cdot \nabla f_s+\frac{q}{m}\left[\vec{E}+\vec{v}\times\vec{B}\right]\cdot \nabla_v f_s=0
\end{equation}
\begin{align*}
    \nabla\cdot \vec{E}&=\rho; ~&\nabla\cdot \vec{B}=0;\\
    ~\nabla\times \vec{E}&=-\frac{\partial\vec{B}}{\partial t}; ~&\nabla\times \vec{B}=\vec{J}+\frac{\partial\vec{B}}{\partial t}
\end{align*}
where $f_i$ represents the number density of each species of particle in phase space,
\begin{equation}
    \rho=\sum_s\int q_sf_s(\vec{x},\vec{v},t) d^3v,
\end{equation}
and
\begin{equation}
    \vec{J}=\sum_s\int q_s\vec{v}f_s(\vec{x},\vec{v},t) d^3v.
\end{equation}

This system can be recast in a form in which the only unitful quantity is the \textit{plasma frequency}, defined as
\begin{equation}
    \label{eq:wpl}
    \omega_{\text{pl},s} = \sqrt{\frac{q_s^2 n_s}{m_s}}
\end{equation}
where $n$ is the number density and $s$ indexes species.
Here, we use units in which $\hbar=c=\epsilon_0=\mu_0=1$. This allows the behavior of plasma at one set of $(q_s, m_s)$ parameters to be generalized to other regions in parameter space at which the plasma takes on the same $\omega_{\text{pl},s}$. 

The Vlasov-Maxwell system is inherently nonlinear; however, when perturbations are small, the equation can be linearized and a dispersion equation computed. 
When a solution to the dispersion relation contains a positive imaginary part, small perturbations grow exponentially, giving rise to an instability. The growth rate of instabilities in the linear regime can be computed analytically in certain simplified settings; however, once the perturbation has grown sufficiently large that  the linear approximation breaks down, 
the dynamics enter the \textit{nonlinear regime}, and numerical simulations are needed to characterize the subsequent evolution. In this regime, the exponential growth back-reacts on the background plasma, leading to a saturation of the instability and, often, the formation of non-linear structures such as collisionless shocks \cite{Bret_2020,Bret_Deutsch_2005,Bret_Gremillet_Dieckmann_2010,dieckmann_bret_2017,Bret_Haggerty_Narayan_2024,Bret_Narayan_2022,Bret_Stockem_etal_2016}.

In the following subsections, we will focus on the linear and nonlinear behavior of one astrophysically-relevant system, namely counter-streaming non-magnetized pair plasmas.

\subsection{Linear regime}
\label{app:linear}
A simple system in which instabilities arise consists of two beams of plasma with comparable densities streaming through one another. Though simple, this system captures the essential features of the astrophysical settings we discuss in Section III of the main text. Two primary instabilities dominate this system. The first is the \textit{two-stream instability} \cite{TS_Original}, in which longitudinal modes grow exponentially as a result of electrostatic interactions. The second is the \textit{Weibel instability} \cite{Weibel_Original}, in which modes transverse to the beam's direction of motion are amplified by electromagnetic interactions with the other beam.

\subsubsection{Two-stream instability}
\label{app:twostream}

The electrostatic instability arises due to the presence of an unstable longitudinal mode in the counter-streaming plasmas. 
The growth rate of perturbations can be calculated in the linear regime via the plasma dispersion relation \cite{TS_Original}

\begin{equation}
    \label{eq:TS_rate}
    \Gamma_\text{TS} \approx \frac{1}{2\sqrt{2}}\omega_\text{pl} 
\end{equation}
with the fastest growing mode having $k_{x,\text{TS}}^{\text{crit}} \approx \frac{\sqrt{3/2}}{v_0}\omega_\text{pl}$, where the subscript TS denotes ``two-stream'' and $v_0$ is the initial relative bulk velocity of the plasma beams. This growth rate has been computed under the assumption that the relative velocity of the beams ($v_0$) is significantly larger than their velocity dispersions (the ``cold limit''). The inclusion of a non-zero temperature only serves to weaken the instability and suppress the growth rate.

The instability saturates when the potential becomes large enough that approaching particles become trapped in the troughs between regions of high potential. This increases the charge density in the trough, raising its potential and reducing the amplitude of the mode.

\subsubsection{Weibel instability}

While the two-stream instability is driven entirely by interactions with the electric field, hence is  ``electrostatic,'' there is an additional electromagnetic instability in the system known as the Weibel instability.\footnote{Additional electromagnetic oblique modes can occur in asymmetric beam plasmas. For beams that are approximately symmetric, as in this work, their growth rates and saturation mechanisms only vary slightly and can be determined numerically. See \cite{Bret_Gremillet_Dieckmann_2010} for further discussion.} 
The growth rate of perturbations can be calculated in the linear regime and is given by \cite{Weibel_Original}
\begin{equation}
    \label{eq:weibel_rate}
    \Gamma_{\text{W}} \approx v_0 \omega_\text{pl}
\end{equation}
with the instability approaching its maximum for modes with $k\gtrsim k_{y,\text{W}}^\text{crit} \approx \omega_\text{pl}$ \cite{dieckmann_bret_2017} in the zero-temperature limit. Note that the Weibel instability grows at a rate suppressed by $v_0$ with respect to the two-stream instability, as is expected given that the two-stream instability arises due to the $q \mathbf{E}$ term in the Vlasov equation while the Weibel instability arises due to the $q \mathbf{v}\times \mathbf{B}$ term.

While the Weibel instability grows more slowly than the two-stream instability at non-relativistic velocities, it still has a significant effect on late-time dynamics. The instability saturates when the magnetic field becomes sufficiently strong such that the gyroradius of particles becomes comparable to the spatial scale of the perturbation. At this stage, deflected particles can no longer penetrate through the magnetic filaments formed during instability growth, and the instability mechanism is halted. Saturation of the Weibel instability generally occurs at a higher electromagnetic field density than the two-stream instability hence, at late times, it is the nonlinear behavior of this instability that dominantly drives the evolution of the plasma.

\subsection{Nonlinear regime}

The growth of the two-stream instability will cause the formation of ``collisionless shocks'' \cite{Livadiotis_2019, Bret_2020,dieckmann_bret_2017,Skoutnev_2019,gary_1993,Bret_Gremillet_Dieckmann_2010,Plasma_Textbook,Bret_Stockem_Fiúza_Álvaro_Ruyer_Narayan_Silva_2013}, shock fronts between the beams supported by collective electromagnetic interactions despite the mean-free-path of an individual particle being much longer than the scale of the shock. The formation of this structure halts when the exponential growth rate becomes comparable to the bounce frequency of the streaming particles \cite{Bret_Gremillet_Dieckmann_2010}
\begin{equation}
\label{eq:TS_trap}
    \Gamma_{TS}^2\approx \omega_{b,E}^2 =\frac{q_\chi E k_{x}}{m_\chi}.
\end{equation}
This process interrupts the bulk motion of the counter-steaming plasma, slowing down the relative bulk velocity while heating up the plasma \cite{Livadiotis_2019}. 

While electrostatic modes initially generate collisionless shocks before dissipating quickly, Weibel filaments will continue to grow and support these shocks on much longer timescales, further contributing to the slowdown of the beams. The Weibel instability saturates when the exponential growth rate becomes comparable to the magnetic trapping frequency \cite{Bret_Gremillet_Dieckmann_2010}
\begin{equation}
    \label{eq:W_trap}
    \Gamma_{W}^2\approx \omega_{b,B}^2 =\frac{q_\chi B v_0 k_{y}}{m_\chi}
\end{equation}
which in general corresponds to a larger electromagnetic field strength than at two-stream saturation.

After saturation, the Weibel filaments will undergo macroscopic instabilities which lead to further turbulence and slowdown. See Appendix (see also references \cite{NH1,NH2,NH3,1D_Weibel,Bret_Weibel,Bret_Gremillet_Dieckmann_2010,Medvedev_2005,Ruyer_Fiuza_2018} therein) for further discussion.

\section{Simulations}
\label{app:simulation}

\subsection{Setup}
Our simulations consist of a 2D3V Cartesian geometry of dimensions $L_x=7\,c/\omega_{pl,B}$ in the longitudinal direction and $L_y=175\,c/\omega_{pl, B}$ in the transverse direction with periodic boundary conditions across all boundaries. Here, $\omega_{pl,B}$ is the plasma frequency corresponding to the Bullet sub-cluster. We choose a grid spacing of $\Delta x = 0.0035\, c/\omega_{pl,B}$ in the longitudinal direction and $\Delta y = 0.18\, c/\omega_{pl,B}$ in the transverse direction to ensure proper resolutions of both two-stream and Weibel instabilities. Each simulation was run for $10^3\,\omega_{pl,B}^{-1}$ with a temporal resolution of $\Delta t =  0.98\Delta x$. Two species were initialized, $\chi$ and $\Bar{\chi}$, with equal and opposite charge-to-mass ratios, 64 total macroparticles per cell were created with second order interpolation schemes for the particle shape function, and bulk velocities $v_{\mathrm{Bullet}}=-n_{\mathrm{Main}} v_0/(n_{\mathrm{Bullet}}+n_{\mathrm{Main}})$ and $v_{Main}=n_{\mathrm{Bullet}} v_0/(n_{\mathrm{Bullet}}+n_{\mathrm{Main}})$ such that the simulation is in the center of momentum frame. Additionally, five multi-pass binomial current filters \cite{VAY20115908} were applied in each dimension after each time step for increased numerical stability. We found this setup to properly resolve all of the relevant length and time scales and produced results with negligible numerical heating.

\begin{figure}[!t]
    \centering
        \includegraphics[width=\textwidth/2]{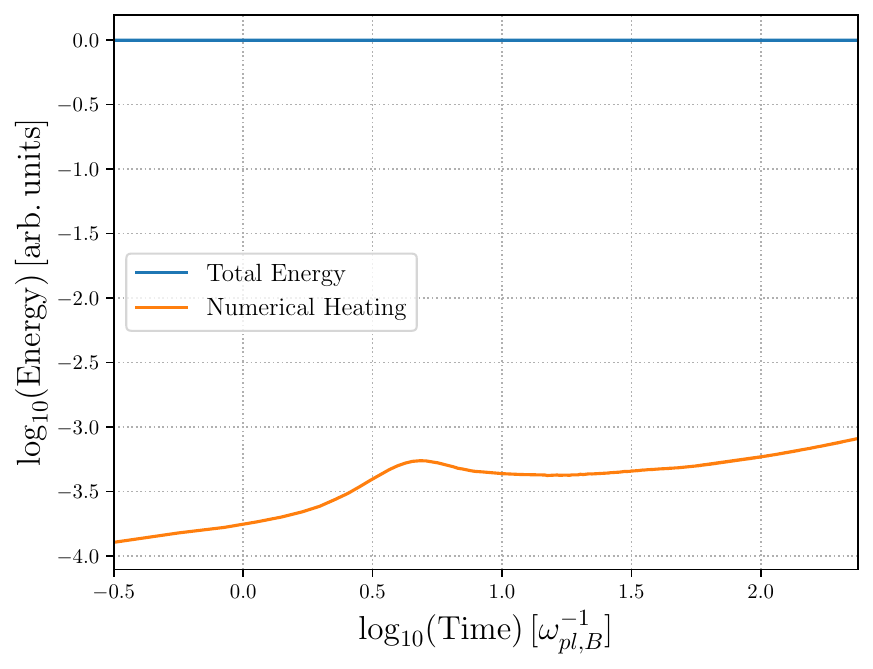}
    \caption{Combined particle and field energy from simulation (blue) and energy lost due to numerical heating (orange).}
    \label{fig:NHeat}
\end{figure}

We undertook several studies to ensure the numerical stability of our simulation. In Fig. \ref{fig:NHeat}, we show the amount of energy increase from numerical heating, a known numerical effect arising from truncation errors and finite-sized time and space domains \cite{NH1,NH2,NH3}. Throughout our simulation, this remained below $0.1\%$ of the total initial kinetic and electromagnetic energy, indicating stability.

\begin{figure}[!t]
    \centering
        \includegraphics[width=\textwidth/2]{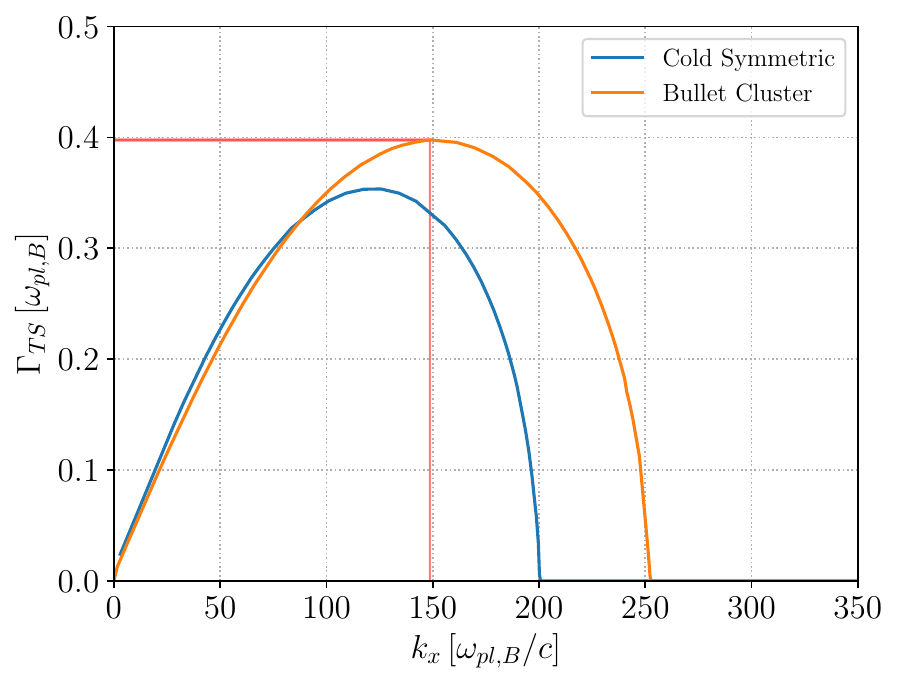}
    \caption{Exponential growth rate of longitudinal two-stream instability in the case of cold symmetric beams (blue) and for the parameters of the Bullet Cluster system (orange). The red lines show the fastest growing mode and corresponding instability rate.}
    \label{fig:TSInst}
\end{figure}

Additionally, we compared the length scales of the two-stream and Weibel instabilities in the asymmetric warm case to ensure that our choice of cell size accurately resolved the relevant instability in each direction. Fig. \ref{fig:TSInst} displays the exponential growth rate for the two-stream instability for the Bullet Cluster system. We find that the maximal growth rate is $\Gamma_{TS}=0.397\,\omega_{pl,B}$ at $k_x=152\,\omega_{pl,B}/c$. This corresponds to a wavelength of $\lambda_{TS}=0.04\,c/\omega_{pl,B}$, which is well above our resolution in the $x$ direction of $\Delta x=0.0035\,c/\omega_{pl,B}$.

\begin{figure}[!b]
    \centering
        \includegraphics[width=\textwidth/2]{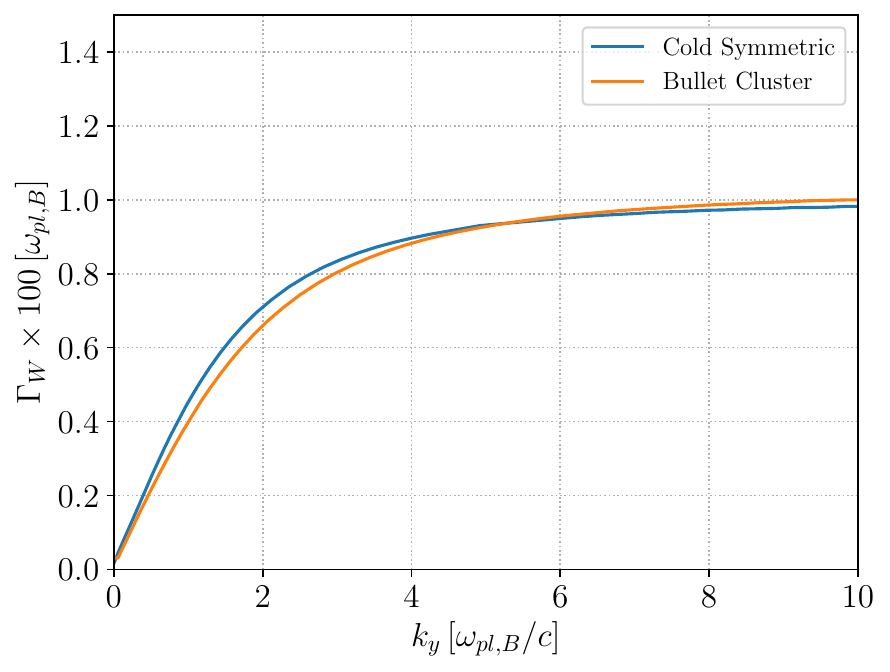}
    \caption{Exponential growth rate of transverse Weibel instability in the case of cold symmetric beams (blue) and for the parameters of the Bullet Cluster system (orange). }
    \label{fig:WInst}
\end{figure}

Similarly, in the transverse direction, Fig. \ref{fig:WInst} shows the exponential growth rate of the Weibel instability. For the Bullet Cluster system, the instability modes reach a maximum of $0.00995\,\omega_{pl,B}$ and display a rapid drop-off for $k<k_{y,\text{W}}^\text{crit}\sim 1 \,\omega_{pl,B}/c$. This corresponds to a maximal wavelength of $\lambda_W= 2\pi \,c/\omega_{pl,B}$, which is also well above our resolution in the $y$ direction of $\Delta y=0.18\,c/\omega_{pl,B}$.

\begin{figure}[!t]
    \centering
    \includegraphics[width=\textwidth/2]{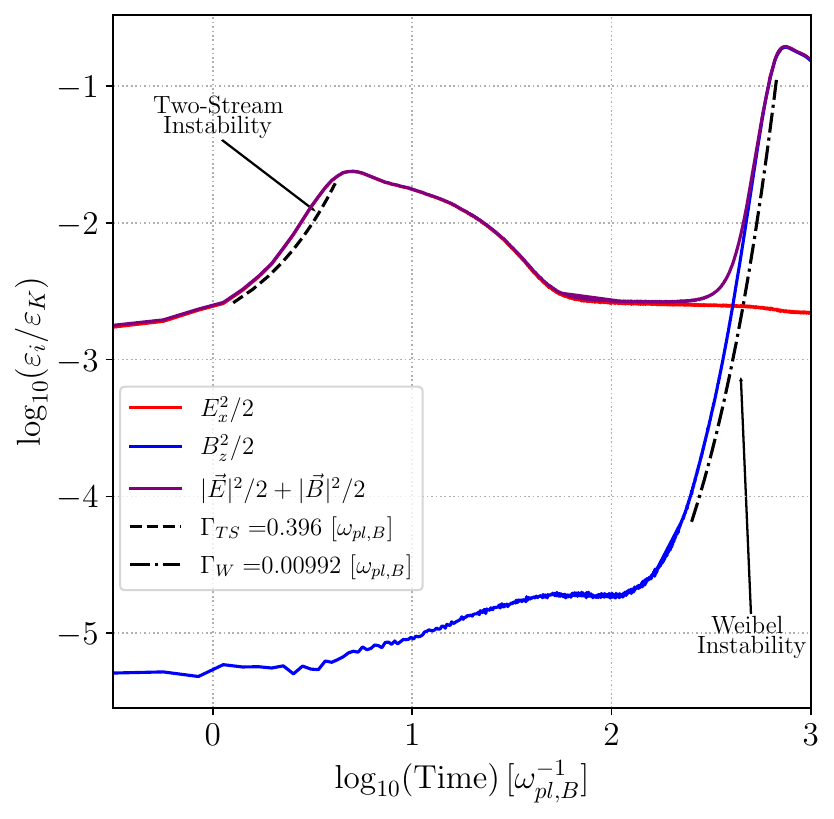}
    \captionsetup{justification=justified}
    \caption{Energy densities from simulation normalized to the initial kinetic energy of the two-beam system. The energy stored in the longitudinal electric fields (red), the transverse magnetic field (blue), and total electromagnetic energy (purple) are shown. Fits to the exponential growth associated with the two-stream and Weibel instabilities are shown as dashed and dashed-dotted lines respectively. }
    \label{fig:Energies}
\end{figure}

\subsection{Growth rates}
Fig. \ref{fig:Energies} shows the energy densities of the longitudinal electric field (red), the transverse magnetic field (blue), and total electromagnetic energy density (purple) from our simulation normalized to the initial kinetic energy of the two-beam system as a function of time. As discussed in Section \ref{app:twostream}, at early times, we expect the two-stream instability to dominate; in Fig. \ref{fig:Energies}, we see that the energy density of the longitudinal electric fields grows exponentially with a growth rate of $\Gamma_{TS}=0.396 \,\omega_{pl,B}$, but saturates rapidly. This growth rate agrees well with numerical estimates. After the electrostatic instability saturates, the Weibel instability continues to grow at 
$0.00992 \,\omega_{pl,B}$, then saturates at $\sim 750 \,\omega_{pl,B}^{-1}$ (Fig. \ref{fig:Energies}).

\subsection{Saturation energies}

From the growth rate in the linear regime, we calculate the approximate saturation energy of the two-stream instability, which yields 
\begin{equation}
    \frac{\varepsilon_{E}}{\varepsilon_K}=\frac{E_{\text{sat}}^2/2}{\frac{1}{2}(4n_0)m_\chi (v_0/2)^2}\approx\frac{1}{96},
    \label{eq:TSSat}
\end{equation}
where $\varepsilon_K$ is the initial kinetic energy density of the plasma. Similarly, we determine the saturation energy for the Weibel instability, which yields
\begin{equation}
    \frac{\varepsilon_{B}}{\varepsilon_K}=\frac{B_{\text{sat}}^2/2}{\frac{1}{2}(4n_0)m_\chi (v_0/2)^2}\approx \left(\frac{v_W}{v_0}\right)^2,
    \label{eq:WSat}
\end{equation}
where $v_W$ is the relative bulk velocity of the plasma during saturation. This value is often less than the initial relative bulk velocity, $v_0$, due to the decrease in bulk kinetic energy during the formation of the electrostatic shock by the two-stream instability.
At regular intervals we sample the velocity distribution of all particles and compute the mean velocity. We find $v_W=0.0014\, c$.
We explicitly measure the saturation levels of the longitudinal electric field and the transverse magnetic field in our simulation in order to compare to these analytic predictions, finding $\varepsilon_E/\varepsilon_K=0.024$ and $\varepsilon_B/\varepsilon_K=0.19$, respectively. We find these values to be within reasonable agreement with Eqs. \ref{eq:TSSat} and Eq. \ref{eq:WSat}.

\subsection{Additional diagnostic plots}
\label{app:Chars}
Here, we show some additional diagnostic plots from our simulation to highlight some of the key features of the evolution of the two-stream and Weibel instabilities. 

\begin{figure*}[htbp]
    \centering
        \includegraphics[width=\linewidth]{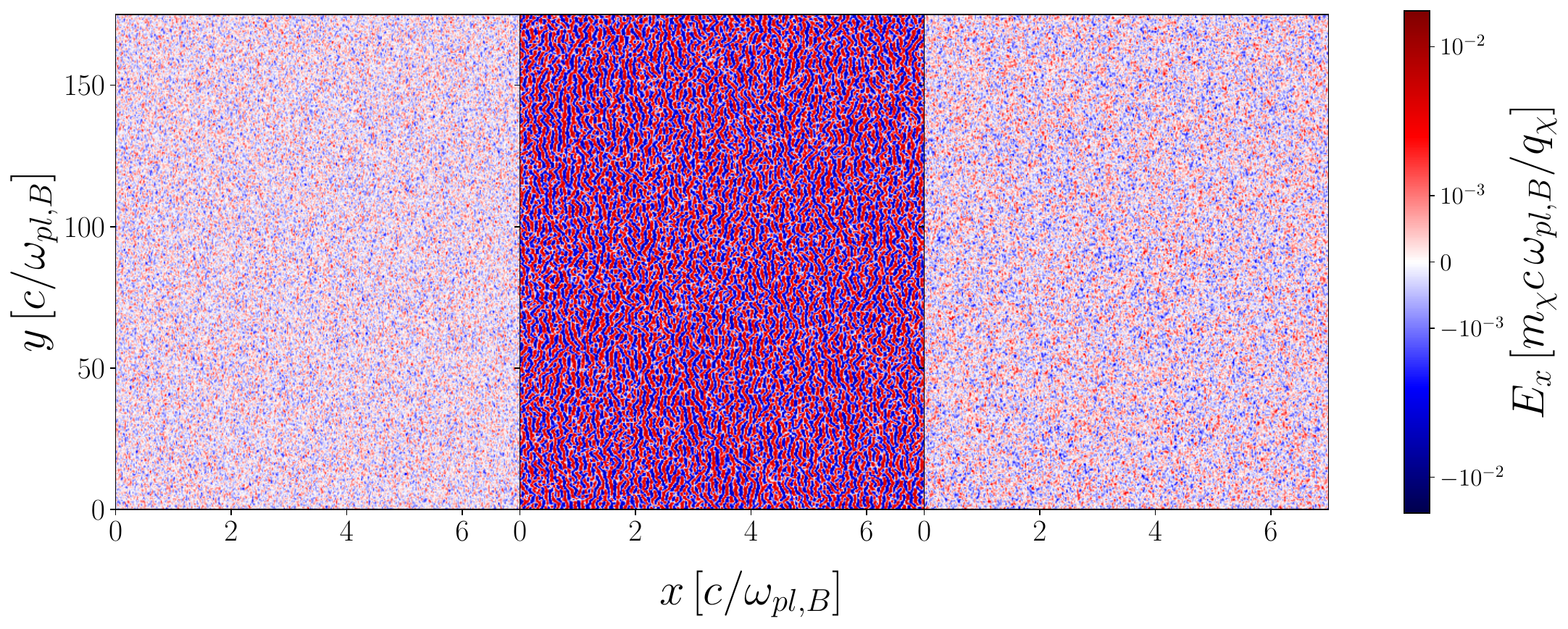}
    \caption{Longitudinal component of the electric field at $t=0.3\,\omega_{pl,B}^{-1}$ (left), $t=5\,\omega_{pl,B}^{-1}$ (middle), and $t=750\,\omega_{pl,B}^{-1}$ (right). }
    \label{fig:Ex}
\end{figure*}

\begin{figure*}[htbp]
    \centering
        \includegraphics[width=\linewidth]{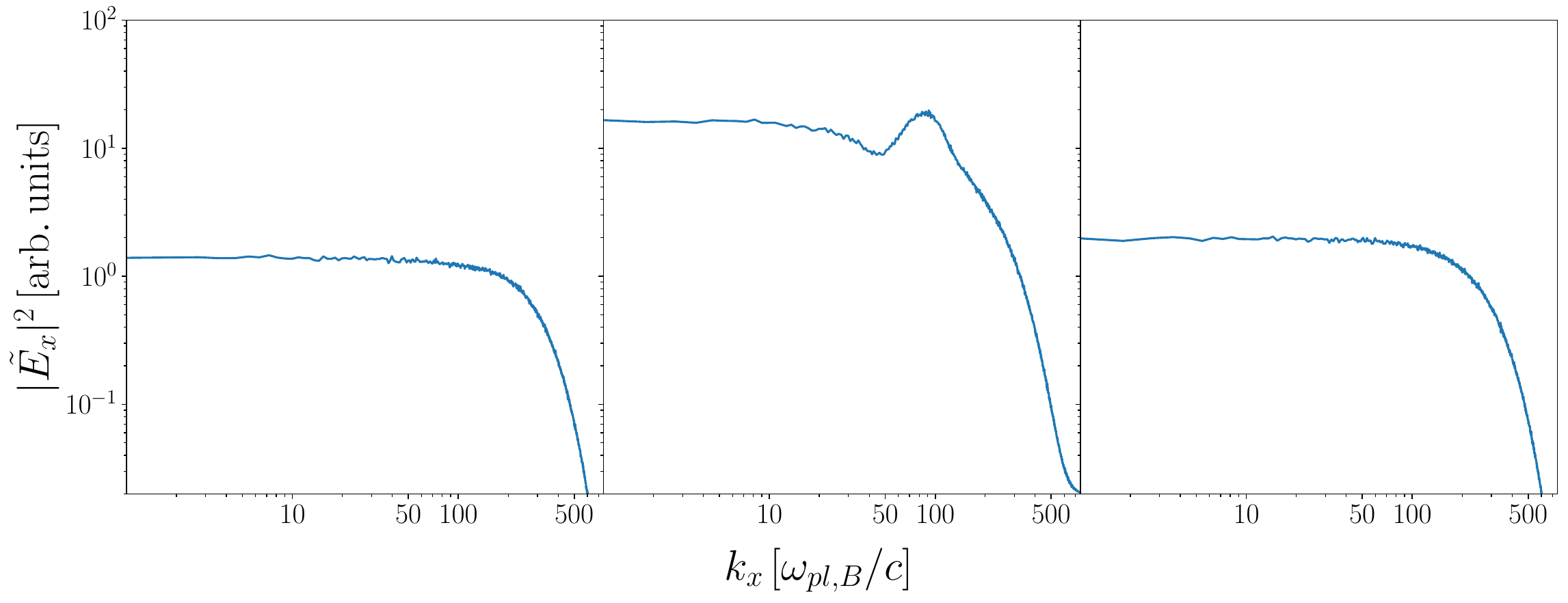}
    \caption{Power spectrum of the longitudinal electric field at $t=0.3\,\omega_{pl,B}^{-1}$ (left), $t=5\,\omega_{pl,B}^{-1}$ (middle), and $t=750\,\omega_{pl,B}^{-1}$ (right).}
    \label{fig:Ex_fft}
\end{figure*}

In Fig. \ref{fig:Ex} (\ref{fig:Ex_fft}), we show a plot of the (Fourier transform of the) electric field along the direction of beam propagation as a function of position for three different times. In the left-most panels, corresponding to $t=0.3\,\omega_{pl,B}^{-1}$, the simulation has properly initialized, and all electromagnetic fields are relatively small and are generated purely by thermal noise effects. At $t=5\,\omega_{pl,B}^{-1}$ (center panels), the two-stream instability has reached saturation, as is indicated by the presence of small vertical strips alternating polarity corresponding to the fastest growing longitudinal mode in Fig. \ref{fig:Ex} and the sharp peak around the critical $k_x$ mode in Fig. \ref{fig:Ex_fft}. At $t=200\,\omega_{pl,B}^{-1}$ (right-most panels), the two-stream instability has dissipated. 

\begin{figure*}[htbp]
    \centering
        \includegraphics[width=\linewidth]{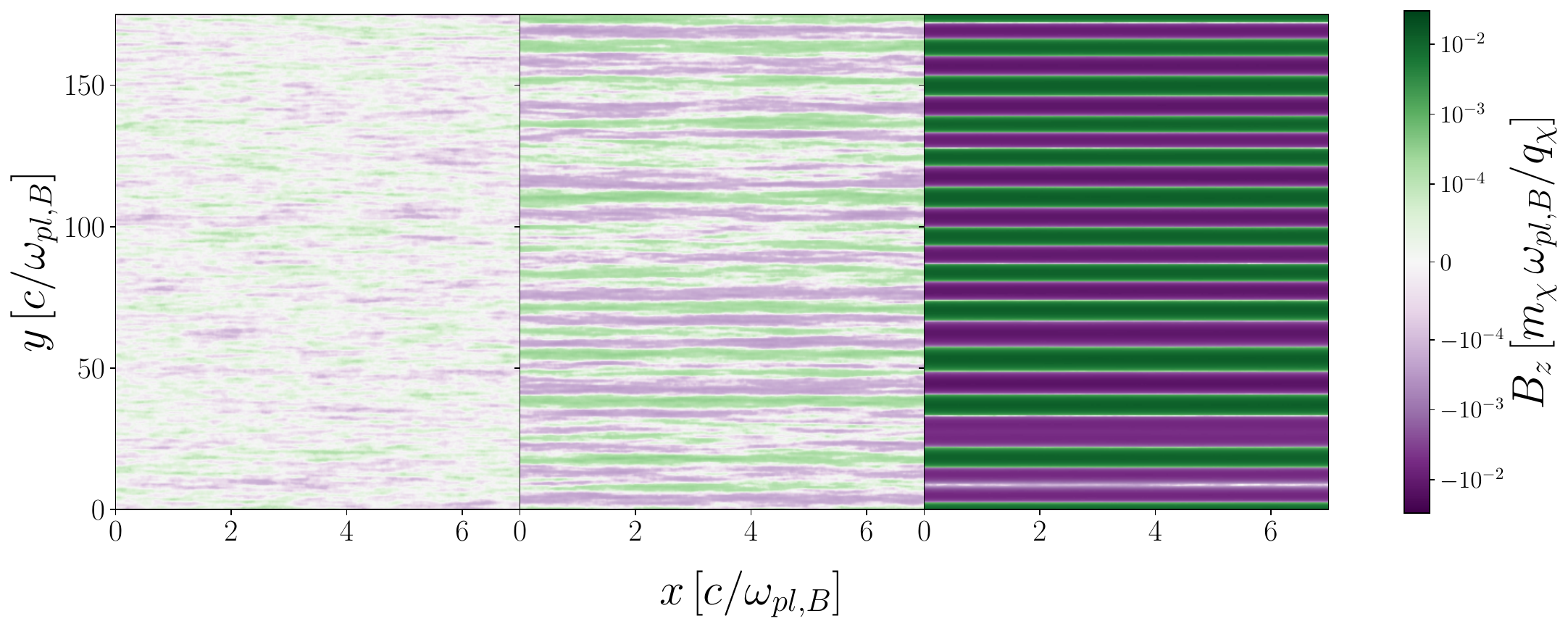}
    \caption{Transverse component of the magnetic field at $t=0.3\,\omega_{pl,B}^{-1}$ (left), $t=200\,\omega_{pl,B}^{-1}$ (middle), and $t=750\,\omega_{pl,B}^{-1}$ (right). }
    \label{fig:Bz}
\end{figure*}

\begin{figure*}[htbp]
    \centering
        \includegraphics[width=\linewidth]{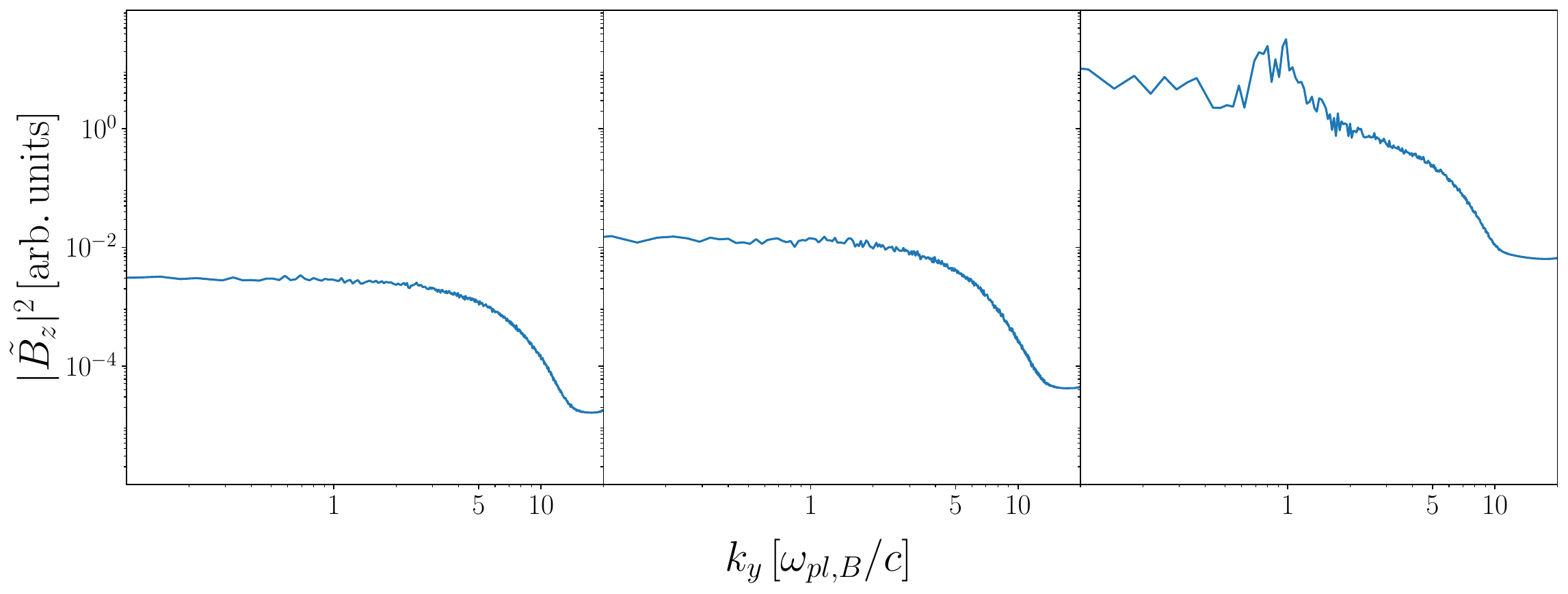}
    \caption{Power spectrum of the transverse magnetic field at $t=0.3\,\omega_{pl,B}^{-1}$ (left), $t=200\,\omega_{pl,B}^{-1}$ (middle), and $t=750\,\omega_{pl,B}^{-1}$ (right).}
    \label{fig:Bz_fft}
\end{figure*}

In Fig. \ref{fig:Bz} (\ref{fig:Bz_fft}), we show a plot of the (Fourier transform of the) magnetic field in the transverse direction as a function of position at the same three times as above.
However, while the two-stream instability has saturated by $t=750\,\omega_{pl,B}^{-1}$, Fig. \ref{fig:Bz} shows that the Weibel filaments are beginning to form and are nearing saturation at this time. Fig. \ref{fig:Bz_fft} shows that the transverse magnetic filaments peak around the critical $k_y$ mode with increasing intensity as the fields reach saturation. 

\subsection{Limitations of 2D3V simulations}
\label{app:3d}

Though our simulation is 2-dimensional, it is still able to conservatively estimate the macrophysical properties of the 3-dimensional Bullet Cluster system. First, we note that the Weibel instability has previously been studied in 1 and 2-dimensional systems \cite{1D_Weibel,Bret_Weibel}, and it has been shown that the growth rates and magnetic field saturation levels agree with analytical models. After Weibel saturation, we expect our simulation to no longer capture the relevant dynamics, as at this time, the Weibel filaments should enter a late-stage merging process that requires the current filaments to move around in 3-dimensional space. Though we cannot capture this effect, it has been previously studied in the literature \cite{Bret_Gremillet_Dieckmann_2010,Medvedev_2005}, and the magnetic field energy is expected to grow approximately linearly with time. These filaments will eventually become unstable to the z-pinch instability, 
which generally works to slow down the flow of particles and isotropize the magnetic field \cite{Ruyer_Fiuza_2018}. Though these late-time phenomena likely only further slow the beams, we chose to set our limit prior to Weibel saturation, long before these merging events are expected to begin, in an effort to be as conservative as possible.
Furthermore, the correspondence of our results with existing literature that simulates both 2D3V and 3D3V systems (see Section IV of main text) provides strong evidence that our results are trustworthy up through Weibel saturation.

\end{document}